\documentclass[prl,superscriptaddress,twocolumn,longbibliography]{revtex4-1}

\usepackage{amsmath}
\usepackage{amssymb}
\usepackage{amsxtra,color}
\usepackage{latexsym}
\usepackage{dsfont}
\usepackage{float}
\usepackage{graphicx}
\usepackage{mathtools}

\graphicspath{ {attoimages/} }   
\usepackage{braket}
\usepackage{tikz}
\usetikzlibrary{arrows}
\usepackage{pgfplots}
\usepgfplotslibrary{groupplots}
\pgfplotsset{compat=1.3}
\usepackage{lipsum}
\usepackage{color}
\DeclareMathOperator{\tr}{tr}

\usepackage[utf8]{inputenc}
\usepackage{MnSymbol}	

\usepackage{hyperref}

\definecolor{MyDarkGreen}{rgb}{0,0.6,0}
\definecolor{MyDarkBlue}{rgb}{0,0,0.8}
\definecolor{MyDarkRed}{rgb}{0.6,0,0.3}

\hypersetup{breaklinks=true, colorlinks=true,plainpages=true, linktocpage=true, linkcolor=MyDarkBlue, citecolor=MyDarkGreen, urlcolor=MyDarkRed, pdfborder={0 0 0},%
pdfauthor={},%
pdfsubject={Research article},
pdftitle={}%
}

\begin{document}

\title{Continuous variable quantum state tomography of photoelectrons}

\author{H. \surname{Laurell}}
\email{hugo.laurell@fysik.lth.se}
\affiliation{Department of Physics, Lund University, Box 118, 22100 Lund, Sweden}
\author{D. \surname{Finkelstein-Shapiro}}
\affiliation{Instituto de Quimica, Universidad Nacional Autonoma de Mexico, Circuito Exterior, Ciudad Universitaria, Alcaldía Coyoacán C.P. 04510, Ciudad de Mexico}
\author{C. \surname{Dittel}}
\affiliation{Physikalisches Institut, Albert-Ludwigs-Universität Freiburg, Hermann-Herder-Straße 3, 79104 Freiburg, Germany}
\affiliation{EUCOR Centre for Quantum Science and Quantum Computing, Albert-Ludwigs-Universität Freiburg, Hermann-Herder-Straße 3, 79104 Freiburg, Germany}
\author{C. \surname{Guo}}
\affiliation{Department of Physics, Lund University, Box 118, 22100 Lund, Sweden}
\author{R. \surname{Demjaha}}
\affiliation{Department of Physics, Lund University, Box 118, 22100 Lund, Sweden}
\author{M. Ammitzb{\"o}ll}
\affiliation{Department of Physics, Lund University, Box 118, 22100 Lund, Sweden}
\author{R. \surname{Weissenbilder}}
\affiliation{Department of Physics, Lund University, Box 118, 22100 Lund, Sweden}
\author{L. \surname{Neori\v{c}i\'c}}
\affiliation{Department of Physics, Lund University, Box 118, 22100 Lund, Sweden}
\author{S. \surname{Luo}}
\affiliation{Department of Physics, Lund University, Box 118, 22100 Lund, Sweden}
\author{M. \surname{Gisselbrecht}}
\affiliation{Department of Physics, Lund University, Box 118, 22100 Lund, Sweden}
\author{C. L. \surname{Arnold}}
\affiliation{Department of Physics, Lund University, Box 118, 22100 Lund, Sweden}
\author{A. \surname{Buchleitner}}
\affiliation{Physikalisches Institut, Albert-Ludwigs-Universität Freiburg, Hermann-Herder-Straße 3, 79104 Freiburg, Germany}
\affiliation{EUCOR Centre for Quantum Science and Quantum Computing, Albert-Ludwigs-Universität Freiburg, Hermann-Herder-Straße 3, 79104 Freiburg, Germany}
\author{T. \surname{Pullerits}}
\affiliation{Chemical Physics and NanoLund, Lund University, Box 124, 22100 Lund, Sweden}
\author{A. \surname{L'Huillier}}
\affiliation{Department of Physics, Lund University, Box 118, 22100 Lund, Sweden}
\author{D. \surname{Busto}}
\email{david.busto@fysik.lth.se}
\affiliation{Department of Physics, Lund University, Box 118, 22100 Lund, Sweden}
\affiliation{Physikalisches Institut, Albert-Ludwigs-Universität Freiburg, Hermann-Herder-Straße 3, 79104 Freiburg, Germany}

\begin{abstract}
\begin{centering}
We propose a continuous variable quantum state tomography protocol of electrons which result from the ionization of atoms or molecules by the absorption of extreme ultraviolet light pulses. Our protocol is benchmarked against a direct calculation of the quantum state of photoelectrons ejected from helium and argon in the vicinity of a Fano resonance. In the latter case, we furthermore distill ion-photoelectron entanglement due to spin-orbit splitting. This opens new routes towards the investigation of quantum coherence and entanglement properties on the ultrafast timescale. 
\end{centering}
 \end{abstract}
\maketitle

 Thanks to the discovery of high order harmonic generation \cite{McPhersonJOSAB1987,FerrayJPB1988}, attosecond light sources were developed, enabling the study of electron dynamics with high temporal resolution \cite{KrauszRMP2009}. By  energy-time uncertainty, such attosecond light pulses, with a central frequency in the extreme ultraviolet (XUV) range, have broad spectral widths. Hence, their interaction with matter usually results in a photoionization process where the ejected electron populates a broad distribution of continuum states. The resulting electronic state may be either pure  or  mixed.

The first experimental methods developed for the characterisation of attosecond pulses relied on the coherence of the photoionization process.  The Reconstruction of Attosecond Beating By Interference of Two photon transitions (RABBIT) \cite{PaulS2001}, as well as attosecond streaking \cite{HentschelNature2001}, were initially invented to characterize the temporal properties of attosecond light pulses. The same techniques were then applied to determine time delays in the photoionization process \cite{KlunderPRL2011,SchultzeScience2010}. More recently RABBIT was used to measure the spectral amplitude and phase of photoelectrons in the vicinity of Fano resonances \cite{KoturNC2016,GrusonScience2016,Busto2018}. 

In general, these characterization methods are readily applicable to pure quantum states. For mixed states of the ejected electrons, which must be described by a density operator rather than by a state vector in Hilbert space, they are unsuitable. Mixed states occur due to several causes, such as decoherence processes or incomplete measurements of entangled particles or degrees of freedom. Decoherence due to interactions with an environment can be neglected on attosecond and few femtosecond time scales. In contrast, strong coupling, e.g., between electronic and nuclear degrees of freedom in molecular systems \cite{Pabst2011,NishiPRA2019,Arnold2020,Vrakking2021} has the potential to induce mixing when only the electron's (or only the ion's) degrees of freedom are interrogated.  

For the characterization of mixed quantum states, the gold standard is \emph{quantum state tomography} (QST) \cite{LvovskyRMP2009}. It aims at reconstructing an unknown state from a series of projective measurements which yield the state density operator -- its most general quantum description -- and is widely used in quantum optics. QST has also been applied successfully in specific instances of multidimensional spectroscopy, where pairs of coherent pulses are used to extract the populations and coherences of the states \cite{YuenZhouPNAS2011,YuenZhouACS2014}. 
QST has only recently been applied in attosecond science. Trains of photoelectrons have been characterized by discrete variable quantum state tomography, using SQUIRRELS (Spectral Quantum Interference for the Regularized Reconstruction of free-Electron States)~\cite{Priebe2017} in the context of electron microscopy, and using Mixed-FROG (Frequency Resolved Optical Gating)  \cite{BourassinBouchet2020} for photoelectrons created by absorption of attosecond pulse trains. Both methods rely on a retrieval algorithm to reconstruct the photoelectron 
quantum state from the measured spectrogram.

Here we propose a robust tomography protocol -- Kvanttillst\aa nds tomogRafi av AttoseKund ElektroNv\aa gpaket (KRAKEN, engl. ``quantum state tomography of attosecond electron wavepackets'') -- that can reconstruct the photoelectron's quantum state without relying on a retrieval algorithm. This method, in contrast to traditional coherent two-dimensional spectroscopy \cite{Hamm2011}, is based on two synchronized narrowband infrared (IR) probe fields at different central frequencies, delayed relative to the extreme ultraviolet (XUV) field, which creates the photoelectron. The measurements consist in recording the photoelectron spectrum as a function of delay and frequency difference of the probe fields. Our method allows the reconstruction of the photoelectron's continuous variable density matrix with, in principle, arbitrary spectral resolution. We numerically demonstrate the KRAKEN protocol for both pure and mixed photoelectron states, in the vicinity of autoionizing resonances in helium and argon, and compare our results 
to direct calculations of the density matrix. In the case of argon we further apply our protocol to quantify the entanglement between ion and photoelectron.
  
\textit{Density matrix formulation of photoionization.} Let us consider the ionization of an atom or molecule by an XUV light pulse, creating an electron in the continuum. We first consider a single angular momentum channel, e.g. $1\text{s}\rightarrow\epsilon \text{p}$, so that the final state is fully described by its energy $\epsilon$. We use the density matrix formalism, which allows a description of mixed photoelectron states. These could arise from experimental imperfections such as partially incoherent XUV light sources, interactions with a fluctuating environment where different electron/ion pairs evolve under slightly different Hamiltonians, or from an incomplete measurement of the degrees of freedom of an entangled state, e.g. between ion and electron.

We concentrate on the latter case, \textit{i.e.} a state which exhibits entanglement between electron and ion, while assuming a fully coherent XUV light pulse and vanishing interactions with the environment.  In this case, the quantum state of the ion {\em and} photoelectron after the ionization is pure, and can be written as a coherent superposition of different orthonormal ion $\ket{j}$ and photoelectron $\ket{\epsilon}$ states,
\begin{align}
\ket{\Psi_\mathrm{IE}}=\sum_{j} \int \mathrm{d}\epsilon~ c_{j}(\epsilon) \ket{j;\epsilon},
\label{eq2}
\end{align}
where $\sum_j \int \mathrm{d}\epsilon~ |c_j(\epsilon)|^2=1$. An experiment that only measures the ionized electron is fully described by the electron quantum state, 
that is, the reduced density matrix ${\rho}_\mathrm{xuv}$ obtained by tracing over the ionic degrees of freedom,
\begin{align}
{\rho}_\mathrm{xuv}&=\mathrm{tr}_\mathrm{I}\left(\ket{\Psi_\mathrm{IE}}\bra{\Psi_\mathrm{IE}} \right)\\
&=\int\ \mathrm{d}\epsilon_1\mathrm{d}\epsilon_2 \sum_j c_j(\epsilon_1)c_j^*(\epsilon_2)\ket{\epsilon_1}\bra{\epsilon_2}, \label{eq:rhoXUVexample}
\end{align}
with matrix elements $\rho_\mathrm{xuv}(\epsilon_1,\epsilon_2)= \sum_j c_j(\epsilon_1)c_j^*(\epsilon_2)$. Its purity is given by
\begin{align}
\mathrm{tr}({\rho}_\mathrm{xuv}^2)&=\int\ \mathrm{d}\epsilon_1\mathrm{d}\epsilon_2\ \left| \sum_j c_j(\epsilon_1)c^*_j(\epsilon_2)\right|^2
\label{eq:purity} 
\end{align}
and is equal to one if and only if photoelectron and ion are in a separable state, i.e., if the coefficients in Eq.~\eqref{eq:rhoXUVexample} factorize, $c_j(\epsilon)=a_j b(\epsilon)$, with $\sum_j |a_j|^2=\int \mathrm{d}\epsilon\, |b(\epsilon)|^2=1$. Otherwise, $\mathrm{tr}({\rho}_\mathrm{xuv}^2)<1$. In the former case, the photoelectron state is pure, in the latter, it is mixed. Therefore, the purity of one party of the pure bipartite state (\ref{eq2}) characterizes both parties' entanglement \cite{Nielsen2002}.

\begin{figure}[t]
	\includegraphics[width = 1\linewidth]{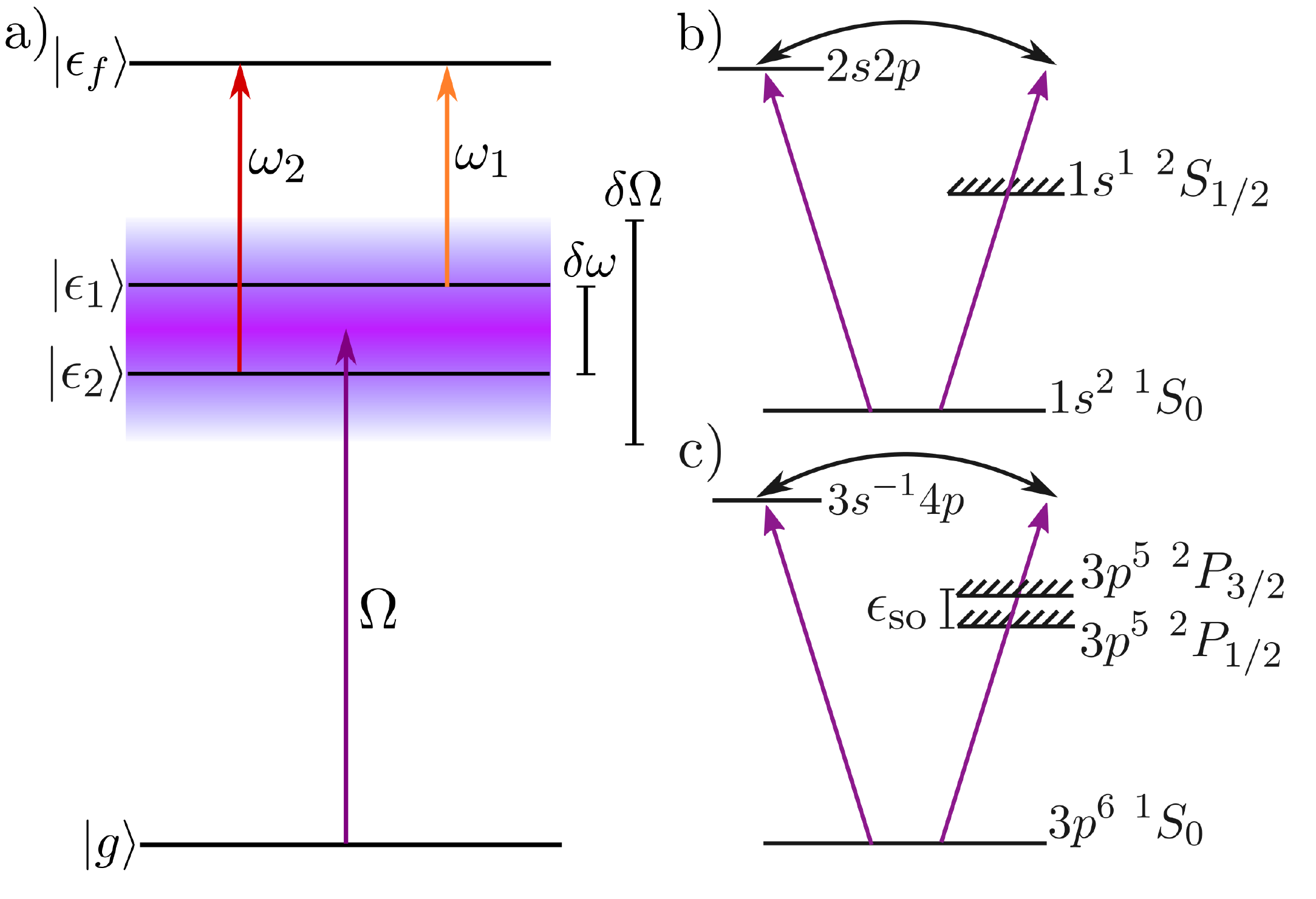}
	\centering
	\caption{(a) Energy diagram of the KRAKEN protocol. The photoelectron is created by absorption of a broadband harmonic with central frequency $\Omega$ and width $\delta\Omega$. The photoelectron's coherence $\rho_\mathrm{xuv}(\epsilon_1,\epsilon_2)$ between the continuum states $\ket{\epsilon_1}$ and $\ket{\epsilon_2}$ is interrogated via simultaneous coherent 
	population transfer from $\ket{\epsilon_1}$ and $\ket{\epsilon_2}$ into the continuum state $\ket{\epsilon_f}$. The population of the latter is modulated by the phase-dependent	interference term of the two amplitudes mediated by the bichromatic IR probe's components $\omega_2$ and $\omega_1$, see Eq.~(\ref{eq8}). $\rho_\mathrm{xuv}(\epsilon_1,\epsilon_2)$ is fully mapped out by scanning $\delta\omega=\omega_2-\omega_1$ over a suitably chosen energy range in the continuum. Panels b) and c) show the energy diagrams for the helium 2s2p (top) and argon 3s$^{-1}$4p (bottom) Fano resonances,	respectively. }
	\label{fig1}
\end{figure}

\textit{KRAKEN protocol.} Without any assumptions on the quantum state of the electron except for being a stationary state, we now show how ${\rho}_\mathrm{xuv}$ can be reconstructed by KRAKEN. As depicted in Fig.~\ref{fig1}(a), we use a broadband XUV light pulse with central frequency $\Omega$ and spectral width $\delta \Omega$ to ionize the target atom. Two synchronised, spectrally narrow probe pulses with central frequencies $\omega_1$ and $\omega_2$ subsequently probe the photoelectron. The absorption or emission of IR photons leads to the formation of additional peaks in the photoelectron spectrum. The use of a bichromatic IR pulse couples different continuum states within the photoelectron's bandwidth and makes them interfere in the final state. 
Measurements of the resulting photoelectron spectra, as a function of the delay between the XUV pulse and the bichromatic IR field, allow us to characterize the coherences 
(i.e., off-diagonal elements) of ${\rho}_\mathrm{xuv}$, between states with energy difference $\delta\omega=\omega_1-\omega_2$. A scan over $\delta\omega$, e.g. by keeping $\omega_1$ constant and scanning $\omega_2$ from $\omega_1$ to $\omega_1+\delta \Omega$, completes the full reconstruction of ${\rho}_\mathrm{xuv}$.

To model the KRAKEN's action upon ${\rho}_\text{xuv}$, we solve the Liouville-von Neumann equation in first order perturbation theory (see Suppl. Mat. (SM) for details). Assuming monochromatic probe fields, we introduce the photoelectron's density operator ${\rho}_\mathrm{xuv+ir}$ after absorbing one photon from the bichromatic IR field (see Fig.~\ref{fig1}(a)). For a given delay $\tau$ between the XUV and IR fields, and frequency difference $\delta\omega$ between the spectral components of the bichromatic IR field, the probability to find a photoelectron with energy $\epsilon_f$ is given by
\begin{equation}
\begin{gathered}
S(\epsilon_f,\tau,\delta\omega)=\bra{\epsilon_f}{\rho}_{\text{xuv+ir}}(\tau,\delta\omega)\ket{\epsilon_f}.\label{eq7}
\end{gathered}
\end{equation}
 $S(\epsilon_f,\tau,\delta\omega)$ can be expressed in terms of the matrix elements of ${\rho}_\text{xuv}$ as
\begin{eqnarray}
\nonumber
S(\epsilon_f,\tau,\delta\omega) &\approx |\mu_{\epsilon_f,\epsilon_1}|^2 \rho_\mathrm{xuv}(\epsilon_1,\epsilon_1)
 + |\mu_{\epsilon_f,\epsilon_2}|^2\rho_\mathrm{xuv}(\epsilon_2,\epsilon_2)\\ \nonumber &+e^{i\delta\omega\tau} \mu_{\epsilon_f,\epsilon_1} \mu^*_{\epsilon_f,\epsilon_2}\rho_\mathrm{xuv}(\epsilon_1,\epsilon_2) \\ 
&+e^{-i\delta\omega\tau}\mu_{\epsilon_f,\epsilon_2} \mu^*_{\epsilon_f,\epsilon_1}\rho_\mathrm{xuv}(\epsilon_2,\epsilon_1),
\label{eq8}
\end{eqnarray}
where $\epsilon_i = \epsilon_f - \hbar\omega_i$ ($i={1,2}$), and $\mu_{\epsilon_f,\epsilon_i}$ are the dipole transition matrix elements between the continuum states $\ket{\epsilon_i}$ and $\ket{\epsilon_f}$ [Fig.~\ref{fig1}(a)].
The first two terms correspond to populations, while the last two terms describe coherences. To extract the coherences, we Fourier transform $S(\epsilon_f,\tau,\delta\omega)$ with respect to $\tau$ and extract the components oscillating at $\pm\delta\omega$, assuming that the dipole transition matrix elements $\mu_{\epsilon_f,\epsilon_i}$ between continuum states 
are constant across $\delta \Omega$.

\textit{Impact of Fano resonances.} As a final preparation for the analysis of our results, we now inspect the impact of the atomic/ionic structure on the photoelectron's spectral density in the continuum. We examine two cases extensively studied using RABBIT \cite{KoturNC2016,GrusonScience2016,Busto2018,Turconi2020,Cirelli2018}, the 2s2p resonance in helium and the 3s$^{-1}$4p in argon as shown in Figure \ref{fig1}(b,c), and
use Fano's description of an autoionizing resonance with energy $\epsilon_r$ and width $\Gamma_r$. In helium, there is only one ionic state, and Eq.~\eqref{eq2} reads \cite{FanoPR1961}
\begin{align}
\ket{\Psi_\mathrm{IE}}= \int \mathrm{d}\epsilon~ E_\text{XUV}(\epsilon)\frac{\Delta(\epsilon)+q}{\Delta(\epsilon)+\text{i}} \ket{^2S_{1/2};\epsilon},
\end{align}
where $\Delta(\epsilon)=2(\epsilon-\epsilon_r)/\Gamma_r$ is the reduced energy,  $E_\text{XUV}(\epsilon)$ the amplitude of the XUV pulse and $q$ the asymmetry parameter proportional to the ratio between the transition matrix elements towards the quasi-bound state at $\epsilon_r$ and  the continuum state at $\epsilon$, respectively.

The case of the 3s$^{-1}$4p resonance in argon is more complicated for two reasons: there are two ionic states $j\in\{^2P_{1/2}, ^2P_{3/2}\}$ separated by an energy $\epsilon_\text{so}$ [see Fig.~\ref{fig1}(c)] and the photoelectron can have two angular momenta  $\lambda\in\{0,2\}$. Consequently, Eq.~\eqref{eq2} generalizes to
 \begin{align} 
\ket{\Psi_\mathrm{IE}}=\sum_{j,\alpha} \int \mathrm{d}\epsilon~ c_{j,\alpha}(\epsilon) \ket{j;\alpha,\epsilon},
\label{eq10}
\end{align}
where $\alpha$ represents the electron quantum numbers other than its energy, and
\begin{align}
c_{j,\alpha}(\epsilon) = E_\text{XUV}(\epsilon)\frac{\Delta_\alpha(\epsilon)+q_\alpha}{\Delta_\alpha(\epsilon)+i}\, .
\label{cj_fano}
\end{align}
This latter expression has to be substituted for the $c_j$ in (\ref{eq:rhoXUVexample}) (with a trivial dependence on $\alpha$ in the case of a helium target), to provide a realistic 
description of ${\rho}_\mathrm{xuv}$.

\begin{figure*}
	\centering
        \includegraphics[width=0.8\linewidth]{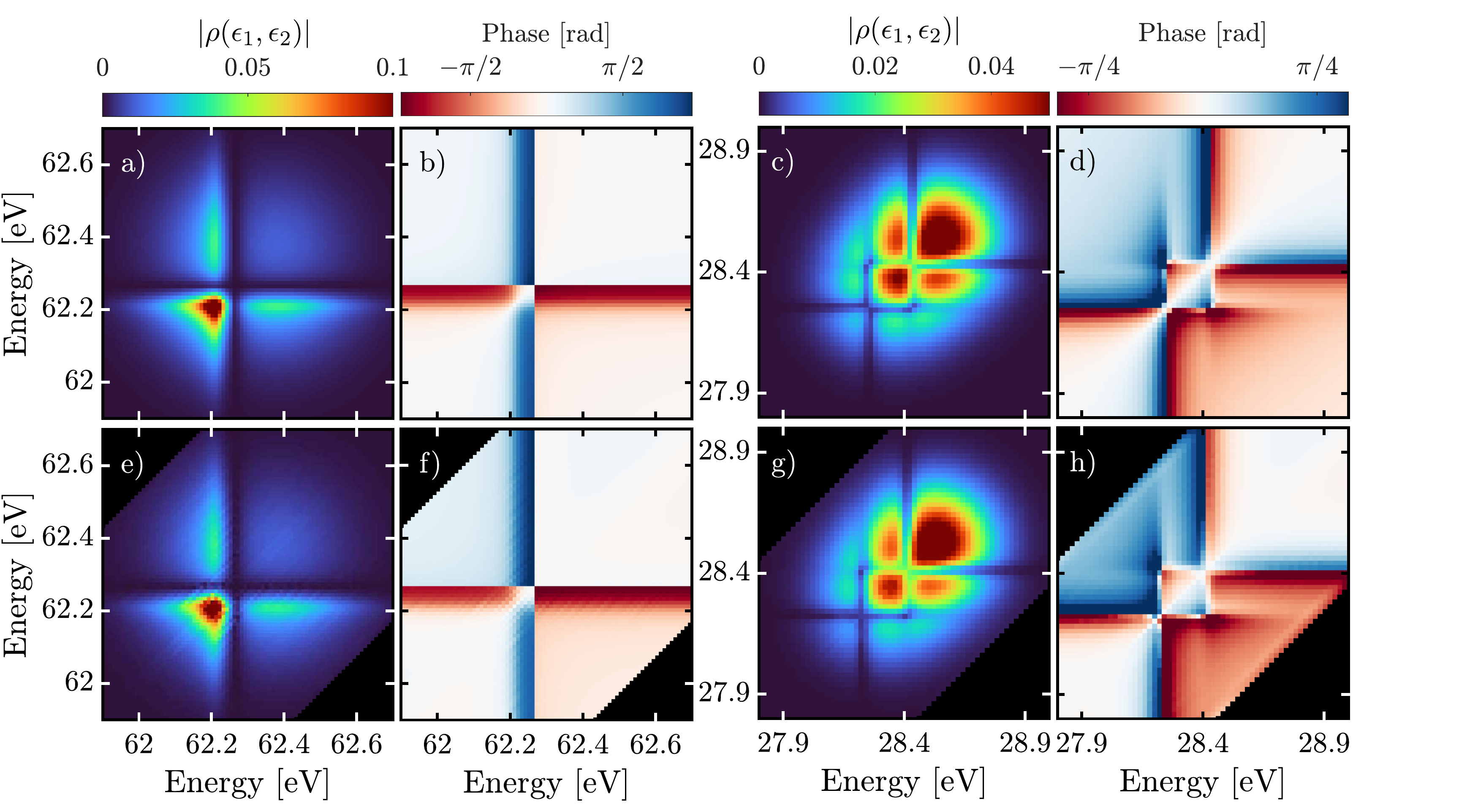} 
        \caption{
        Intensity ($|\rho_\mathrm{xuv}(\epsilon_1,\epsilon_2)|$) and phase profiles of the photoelectron's (reduced) state $\rho_\mathrm{xuv}$ as obtained by direct evaluation 
        of (\ref{eq:rhoXUVexample}) with (\ref{cj_fano}) (upper panels), compared to the output of a simulation of the KRAKEN protocol, based on the observable (\ref{eq8}) (bottom).
        The (double) cross-like structures featured by both, $|\rho_\mathrm{xuv}(\epsilon_1,\epsilon_2)|$ and the phase, are manifestations of the Fano profiles modulating the
        $c_j$ in (\ref{eq2},\ref{eq:rhoXUVexample},\ref{eq10}) according to (\ref{cj_fano}), induced by the $\mathrm{2s2p}$ ($3s^{-1}4p$) resonance of helium (argon). Black areas 
        in (e-h) were not mapped out in our KRAKEN simulation.}
         \label{fig2} 
\end{figure*}

Figs.~\ref{fig2}(a-d) show the modulus and the phase portrait of ${\rho}_\mathrm{xuv}$ as given by (\ref{eq:rhoXUVexample}) with (\ref{cj_fano}), in helium (a,b) or argon (c,d). While the helium photoelectron's state has roughly circular support in the energy plane, the argon state lives on a rather elliptic domain, indicative of coherences between energetically distant states being suppressed. Furthermore, the Fano structure induced by the autoionizing state decaying towards one (helium -- a,b) or two (argon -- c,d) non degenerate ionic states is
clearly reflected by the sharp, cross-like structure(s) both in the modulus as well as in the phase of ${\rho}_\mathrm{xuv}$.

The above portraits of ${\rho}_\mathrm{xuv}$ are now to be compared to the spectrograms produced by KRAKEN -- which are calculated using an analytical model for resonant two-photon transitions \cite{JimenezGalan2016}\footnote{Note that the assumption that the continuum-continuum matrix elements are constant across $\delta\Omega$ is still valid in the presence of the Fano resonance in the intermediate state \cite{JimenezGalan2016}.}  (further details in the SM). We choose an XUV field with a central photon energy of 60.25 eV in helium (26.70 eV in argon). In both cases, the XUV bandwidth is 0.25 eV (full width at half maximum), and the IR bandwidth is set to 1.5~nm in helium (3~nm in argon). We vary the pulse delay between -200 fs to 200 fs in helium (-400 fs to 400 fs in argon) with a step size of 3 fs, and the frequency difference $\delta\omega$ is varied in steps of 0.01~eV.

We calculate spectrograms for a broad range of $\delta\omega$, to map out the 
subdiagonals of the reconstructed density matrix ${\rho}_\mathrm{xuv}^\text{rec}$,and smoothen the result by a moving local average (further details on the reconstruction procedure in the SM). The results are shown in Figs.~\ref{fig2} (e,f) for helium, and (g,h) for argon, respectively, and, by visual inspection, agree well with the direct evaluation of (\ref{eq:rhoXUVexample},\ref{cj_fano}) in panels (a-d). 
To validate this impression, we evaluate the fidelity of ${\rho}_\mathrm{xuv}$ 
to ${\rho}_\mathrm{xuv}^\mathrm{rec}$, according to \cite{Nielsen2002} 
\begin{align}
F({\rho}_\mathrm{xuv},{\rho}_\mathrm{xuv}^\mathrm{rec} )=\mathrm{tr}\left[\left(  {\rho}_\mathrm{xuv}^{1/2}   {\rho}_\mathrm{xuv}^\mathrm{rec}{\rho}_\mathrm{xuv}^{1/2} \right)^{1/2}\right]\, .
\end{align} 
We find $F=0.9997$ for helium ($0.9718$ for argon). The minor deviation from unity is mainly caused by errors stemming from finite bandwidths of the two IR probe pulses.  
The presence of two intermediate angular momentum channels explains the slightly larger mismatch for argon.

\begin{figure}[h]
	\centering
        \includegraphics[width=1\linewidth]{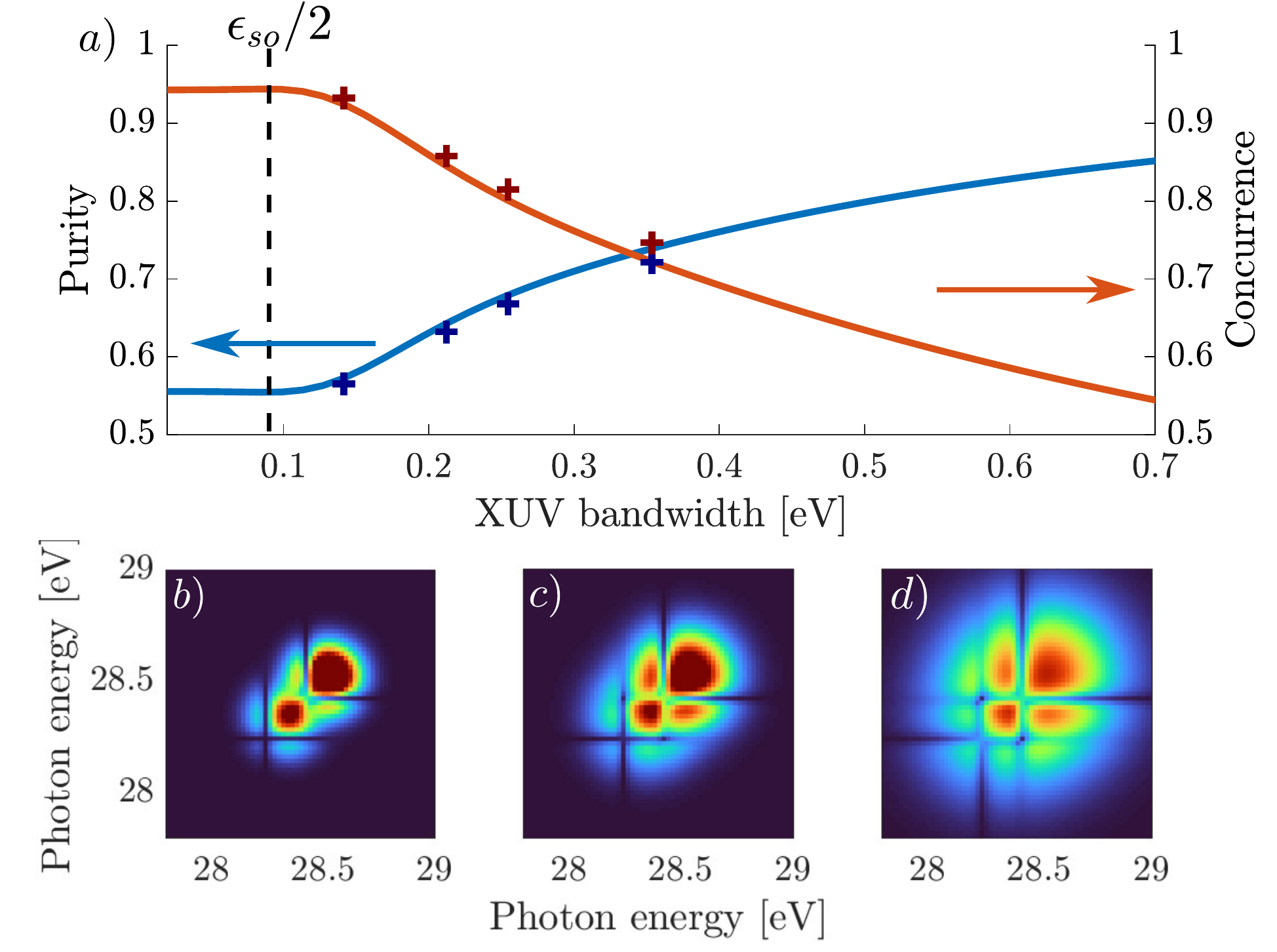} 
        \caption{Entanglement between ion and photoelectron in argon. In the top panel (a), blue (red) solid lines show the directly calculated purity 
        $\mathrm{tr}({\rho}_\mathrm{xuv}^2)$ (concurrence $\mathcal C$) as a function of the XUV bandwidth, and blue (red) crosses correspond to the 
        values reconstructed via the KRAKEN protocol. The vertical black dashed line indicates half of the spin-orbit splitting energy $\epsilon_\mathrm{so}$ for the $3s^{-1}4p$ Fano resonance in argon, such that bandwidths smaller than $\epsilon_\mathrm{so}/2$ resolve the spin-orbit states. On the bottom, the modulus of the photoelectron state $|\rho_\mathrm{xuv}(\epsilon_1,\epsilon_2)|$ as given by (\ref{eq:rhoXUVexample},\ref{cj_fano}) is shown for XUV bandwidths $\delta\Omega=0.14\,\mathrm{eV}$  (b), $\delta\Omega=0.21\,\mathrm{eV}$  (c), and $\delta\Omega=0.35\,\mathrm{eV}$ (d).}
         \label{fig4} 
\end{figure}

\textit{Ion-electron entanglement.} Finally, we investigate how to control and measure the degree of electron-ion entanglement in argon, by varying the bandwidth of the XUV radiation. To this end, recall that for pure bipartite states~\eqref{eq2} of electron and ion, the purity $\mathrm{tr}({\rho}_\mathrm{xuv}^2)$ of the 
photoelectron (reduced) state ${\rho}_\mathrm{xuv}$ {\em decreases} with {\em increasing} entanglement (this simply expresses the loss of information upon disregarding the ionic degree of freedom). The purity is related to the concurrence, a measure of entanglement, through \cite{MintertPR2005}
\begin{align}
\mathcal{C}=\sqrt{2[1-\mathrm{tr}({\rho}_\mathrm{xuv}^2)]}\, .
\end{align}
Both quantities are plotted in Fig.~\ref{fig4}(a) as function of the XUV bandwidth. Solid lines result from our direct evaluation of (\ref{eq:rhoXUVexample}), while crosses are obtained from our simulation of the KRAKEN protocol. We see that entanglement (purity) decreases (increases) with increasing XUV bandwidth. For XUV bandwidths smaller than the spin-orbit splitting $\delta\Omega < \epsilon_\text{so}$, the two spin-orbit components $^2P_{1/2}$ and $^2P_{3/2}$ of the ion are spectrally fully resolved, such that there is maximal entanglement between ion and photoelectron [left of the dashed line in Fig.~\ref{fig4}(a)], and the detection of an electron with a given kinetic energy allows us to determine the ionic state with certainty. On the other hand, when the XUV bandwidth is increased, the two photoelectron spectra associated with the spin-orbit components start to overlap. Hence, the overlap region reveals no information about the ionic state. As a result, the degree of entanglement between photoelectron and ion decreases. In the limit $\delta\Omega \gg \epsilon_\text{so}$, the two photoelectron spectra fully overlap, such that the energy of the electron does not reveal any information on the ionic state and there is no entanglement.

To be more precise, in our model the reduced density matrix of the photoelectron can be expressed as \cite{Turconi2020} 
\begin{equation}
{\rho}_\text{xuv}= \tfrac{1}{3}{\rho}_{1/2}+\tfrac{2}{3}{\rho}_{3/2},
\end{equation}
where  ${\rho}_{1/2}$ and ${\rho}_{3/2}$ are identical, pure density matrices shifted in energy by $\epsilon_\text{so}$, according to $\rho_{1/2}(\epsilon_1,\epsilon_2)=\rho_{3/2}(\epsilon_1-\epsilon_\text{so},\epsilon_2-\epsilon_\text{so})$. 
$\mathrm{tr}({\rho}_{1/2}^2)=\mathrm{tr}({\rho}_{3/2}^2)=1$ implies 
\begin{align}\label{eq16}
\mathrm{tr}({\rho}_\mathrm{xuv}^2)=\tfrac{5}{9} + \tfrac{4}{9}\tr({\rho}_{1/2}{\rho}_{3/2}).
\end{align}
For non-overlapping photoelectron spectra, we have $\mathrm{tr}({\rho}_{1/2} {\rho}_{3/2})=0$, and the purity in~\eqref{eq16} becomes minimal, $\mathrm{tr}({\rho}_\mathrm{xuv})=5/9$, indicating maximal entanglement, $\mathcal{C}=2\sqrt{2}/3\approx 0.94$. On the other hand, for fully overlapping spectra, $\mathrm{tr}({\rho}_{1/2} {\rho}_{3/2})=1$, the purity reaches its maximum, $\mathrm{tr}({\rho}_\mathrm{xuv}^2)=1$, so that there is no entanglement between ion and photoelectron, $\mathcal{C}=0$.

The decrease of entanglement with increasing XUV bandwidth is also visible in the amplitude of the density matrix elements. As shown in Figs.~\ref{fig4}(b)-(d), when the XUV bandwidth is increased, the support of ${\rho}_\mathrm{xuv}$ expands into the energy plane, since the two spin-orbit states have larger overlap. An expanding support of ${\rho}_\mathrm{xuv}$  is tantamount of larger coherences inscribed into the reduced state of the photoelectron, which amounts to decreasing entanglement with the mother ion.

In conclusion, we presented a continuous variable quantum state tomography protocol -- KRAKEN -- to characterize photoelectrons created by absorption of XUV pulses. We tested KRAKEN in the vicinity of Fano resonances of helium and argon. In both cases, a comparison of numerical simulations of KRAKEN output with direct calculations of the density matrix showed excellent agreement. For the Fano resonance in argon, we further showed how to control and measure the degree of entanglement between the ion and the photoelectron by varying the bandwidth of the XUV pulse. Let us stress that the proposed tomography protocol is not restricted to Fano resonances. It is generally applicable to 
photoelectrons created by absorption of high-order harmonics or free electron laser pulses of femtosecond or attosecond duration. While we here demonstrated that KRAKEN can be used to study entanglement properties between ion and photoelectron due to spin-orbit interactions, it can likewise be used to investigate other bipartite entanglement processes, e.g. regarding electron-ion entanglement in vibrating molecules \cite{Vrakking2021,KollPRL2022}. This opens up new routes for uncovering the role of entanglement in atomic and molecular physics \cite{TichyJPB2011}. \\

 This work was supported by the Swedish Research Council, the European Research Council and the Knut and Alice Wallenberg Foundation. D.F.S acknowledges funding from PAPIIT No. IA202821. C.D. acknowledges financial support by the Georg H. Endress foundation. A.L. is partly supported by the Wallenberg Center for Quantum Technologies (WACQT), funded by the Alice and Wallenberg Foundation.

\bibliography{Ref_lib}

\end{document}